\documentclass[%
 aip,
 amsmath,amssymb,
reprint,%
]{revtex4-1}

\usepackage{graphicx}
\usepackage{dcolumn}
\usepackage{bm}

\usepackage[utf8]{inputenc}
\usepackage[T1]{fontenc}
\usepackage{mathptmx}
\usepackage{etoolbox}
\usepackage{upgreek}

\makeatletter
\def\@email#1#2{%
 \endgroup
 \patchcmd{\titleblock@produce}
  {\frontmatter@RRAPformat}
  {\frontmatter@RRAPformat{\produce@RRAP{*#1\href{mailto:#2}{#2}}}\frontmatter@RRAPformat}
  {}{}
}%
\makeatother
\begin{document}

\preprint{AIP/123-QED}

\title[Experimental and Numerical Study of Acoustic Streaming in Mid-Air Phased Arrays]{Experimental and Numerical Study of Acoustic Streaming in Mid-Air Phased Arrays}
\author{Christopher Stone}
 \affiliation{School of Electrical, Electronic and Mechanical Engineering, University of Bristol, BS8 1TR Bristol, United Kingdom}
\author{Yusuke Koroyasu}%
\affiliation{Graduate School of Comprehensive Human Sciences, University of Tsukuba, Tsukuba, 305-8550, Japan}%
\author{Yoichi Ochiai}%
\affiliation{Pixie Dust Technologies, Inc, Tokyo, 104-0028, Japan}%
\affiliation{Institute of Library, Information and Media Science, University of Tsukuba, Kasuga Campus Kasuga 1-2, Tsukuba, 305-8550, Ibaraki, Japan}%
\affiliation{R\&D Center for Digital Nature, University of Tsukuba, Kasuga Campus Kasuga 1-2, Tsukuba, 305-8550, Ibaraki, Japan}%
\author{Akiko Kaneko}%
\affiliation{Institute of Systems and Information Engineering, University of Tsukuba, Tennodai 1-1, Tsukuba, 305-0006, Ibaraki, Japan}%
\author{Bruce W. Drinkwater}%
 \affiliation{School of Electrical, Electronic and Mechanical Engineering, University of Bristol, BS8 1TR Bristol, United Kingdom}
\author{Tatsuki Fushimi}
\email{tatsuki@levitation.engineer}
\affiliation{Institute of Library, Information and Media Science, University of Tsukuba, Kasuga Campus Kasuga 1-2, Tsukuba, 305-8550, Ibaraki, Japan}%
\affiliation{R\&D Center for Digital Nature, University of Tsukuba, Kasuga Campus Kasuga 1-2, Tsukuba, 305-8550, Ibaraki, Japan}%
\date{\today}

\begin{abstract}
Mid-air acoustic streaming, where ultrasound induces steady fluid motion, could significantly affect the perception of haptic sensations, stability of levitation systems, and enable controlled transfer of odours (smells) through air by directing volatile compounds to specific locations. Despite its importance, the streaming behavior in airborne phased-array transducers remains poorly understood. Here, we use particle image velocimetry and numerical simulations to investigate streaming dynamics in single- and multi-focus acoustic fields. Experimental measurements reveal streaming velocities exceeding $0.4~\text{m/s}$ in single-focus configurations and up to $0.3~\text{m/s}$ in multi-focus setups, with distinct grating lobe-induced lateral jets. While multi-physics finite-element models effectively capture central streaming, they exhibit subtle differences and perform poorly in capturing flow in the side lobes. These findings provide valuable insights into the interplay between acoustic field design and streaming dynamics, offering guidance for optimizing ultrasonic technologies in haptics and levitation applications.
\end{abstract}

\maketitle

\section{Introduction}
\label{sec:intro}

Acoustic streaming has been extensively investigated in liquid media and microchannels, where ultrasound-induced streaming reveals intricate fluid dynamics~\cite{muller2014numerical, das2020acoustic}. These studies contributed significantly to understanding acoustic streaming in confined liquid geometries and its applications in particle manipulation, fluid mixing, and lab-on-chip technologies~\cite{karlsen2015forces, wiklund2012acoustofluidics, hou2025fluid}.

The phenomenon of mid-air acoustic streaming stems from early work by Hasegawa et al.\cite{Hasegawa2017}, and many follow-up studies have explored applications such as fog \cite{norasikin2019sonicspray}, odor displays~\cite{hasegawa2017interactive} and localized cooling sensations~\cite{Nakajima2021}. Furthermore, acoustic streaming could affect the performance of technologies such as volumetric displays~\cite{fushimi2019acoustophoretic,hirayama2019volumetric} and digital microfluidics~\cite{koroyasu2023microfluidic}. While streaming velocity measurements using mechanical~\cite{norasikin2019sonicspray} and hot-wire anemometers~\cite{Hasegawa2017,Hasegawa2019} have been reported, these methods are prone to inaccuracies from nonlinear acoustic field interactions. Particle image velocimetry (PIV) offers a more direct and reliable method for capturing streaming velocity fields, as demonstrated in measurements of flows driven by Langevin horns and focused transducer arrays~\cite{Stone2024}.

Here, we apply PIV to characterize airborne streaming flows induced by flat phased array transducers (PATs), addressing a critical gap in understanding these systems. PATs can shape and steer acoustic fields to enable advanced airborne applications. This study systematically explores focal distances, beam geometries, and multi-focus configurations, comparing experimental results with numerical predictions under different attenuation assumptions. The results provide new insights into the design of high-power ultrasonic systems, particularly, the influence of focal distance on streaming fields, impact of grating lobes, and transient responses, ultimately offering a comprehensive framework for optimizing ultrasonic applications in haptics, odor displays, and levitation.
\section{Methods}
\label{sec:methods}

\subsection{Experimental setup}
\label{subsec:experimentalsetup}
A 16$\times$16 array of open-type air coupled transducers (MA40S4S, Murata, Japan) based on OpenMPD design was used in the experiment (see Montano-Murillo et al.~\cite{montano2023openmpd} for details on the array). A MOSFET driver (MIC4127YME) was used to drive the array at a frequency of $f=40$~kHz. For PIV measurements, the array was positioned at the bottom of a $1\times 1 \times 1.2$~m enclosed chamber filled with smoke particles (KINCHO Mosquito Coil, Japan Insecticide Mfg. Co.). A 5W green laser was used to illuminate the particles, with the laser plane positioned at $y=0$~mm (that is in the $xz-$plane). A Sony $\alpha 9$ camera (positioned 560 mm from the plane) with a Sony 16-35 mm lens was used to capture the particle displacements at a frame rate of 240 Hz, capturing 1000 frames per case. PIVlab in MATLAB\cite{stamhuis2014pivlab} was used to process the data and calculate the streaming velocity fields.

\subsection{Huygens' Principle Model for the Pressure Field}
The acoustic pressure field was modeled using a 3D Huygens' principle model approach. The total pressure field \(p\) was computed by superimposing contributions from all transducers:
\[
p = \sum_{n}^N \frac{P_{\text{scale}}}{R_n} D(k, \theta_n) e^{j(kR_n + \phi_n)},
\]
where \(P_{\text{scale}} = 0.179 V_{in}\) is the transducer power based on the input voltage \(V_{in}\), \(R_n\) is the distance from the transducer to the field point, \(D(k, \theta)\) is the directivity function (see Supplementary Material), and \(k = 2\pi/\lambda\) is the wavenumber. 

The phase shifts \(\phi_n\) were calculated to generate specific beam types. For focused beams, \(\phi_n\) ensured constructive interference at the focal point, while for Bessel beams, the phases were derived based on the cone angle and transducer geometry. Multi-focus fields were calculated using the iterative backpropagation (IBP) algorithm, which iteratively adjusts transducer pressures to achieve specified amplitudes at multiple focal points. The IBP method was implemented as described in Marzo \& Drinkwater\cite{marzo2019holographic}. Detailed derivations of the phase calculations for focused, Bessel, and multi-focus fields are provided in the Supplementary Material.

\subsection{Acoustic Streaming Force}
The acoustic particle velocity was computed as:
$\textbf{v}_1 = \frac{1}{j \omega \rho} \nabla p$,
where \(\omega = 2\pi f\), \(f\) is the frequency, and \(\rho\) is the air density. Using this velocity, the acoustic intensity vector was calculated as:
$\textbf{I} = \frac{1}{2} \Re(p \cdot \textbf{v}_1^*),
$
where \(\textbf{v}_1^*\) denotes the complex conjugate of the particle velocity. The streaming force per unit volume was then given by:
\[
\frac{\textbf{F}_s}{dV} = \frac{2 \alpha \textbf{I}}{c},
\]
where \(c\) is the speed of sound and \(\alpha\) is the attenuation coefficient. The methods used to calculate \(\alpha\) are described below.

\subsection{Attenuation Coefficients}
Two attenuation mechanisms were incorporated into the model: thermoviscous and atmospheric. 
\begin{figure*}[t]
    \centering
    \includegraphics[width=0.9\textwidth]{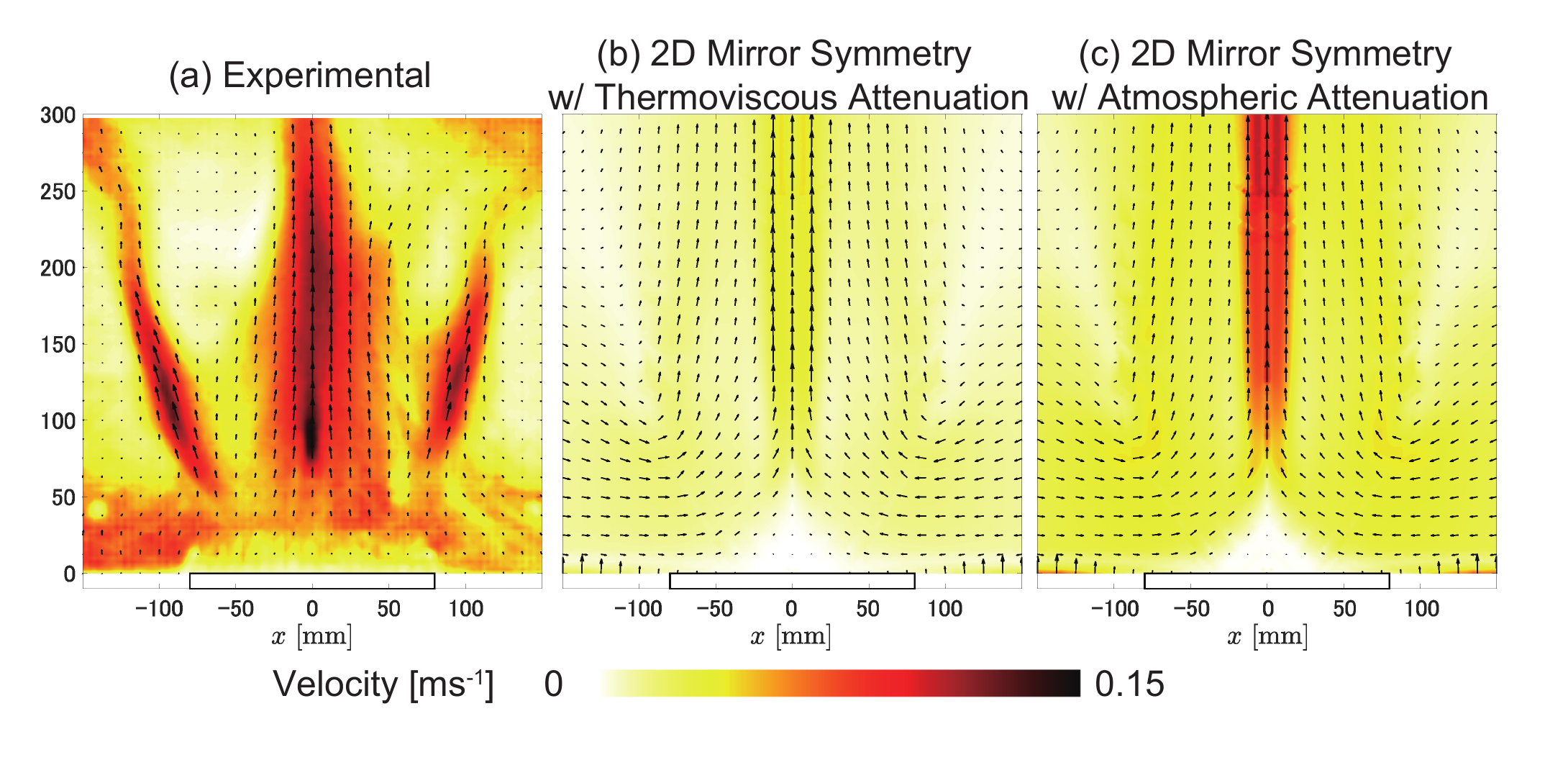}
    \caption{Comparison of experimental (a) and numerical (b–c) streaming velocity fields obtained under different attenuation models. Data were acquired using a focused beam at (0, 0, 80) mm, 5V. Experimental measurements (a) reveal pronounced grating lobe jets alongside a focused central jet. The thermoviscous model (b) significantly underestimates both the central flow and the grating lobe jets. The atmospheric model (c) shows improvement but still underestimates the grating lobe jets.}
    \label{fig:quiver_experiment_numerical}
\end{figure*}
The thermoviscous attenuation coefficient was calculated using the Stokes-Kirchhoff relation\cite{hu2015sound}:
\[
\alpha_{\text{thermoviscous}} = \frac{\omega^2}{2 \rho c^3} \left(\frac{4 \mu}{3} + \mu_B + (\gamma - 1)\frac{k_{\text{cond}}}{C_p} \right),
\]
where \(\mu\) is the dynamic viscosity, \(\mu_B\) is the bulk viscosity, \(\gamma\) is the ratio of specific heats, \(k_{\text{cond}}\) is the thermal conductivity, and \(C_p\) is the specific heat capacity at constant pressure. A detailed derivation of this equation is included in the Supplementary Material.

The atmospheric attenuation coefficient was given by\cite{bass1995atmospheric}:
\[
\alpha_{\text{atmospheric}} = f^2 \left( B_1 \cdot \frac{f_{rN}}{f_{rN}^2 + f^2} + B_2 \cdot \frac{f_{rO}}{f_{rO}^2 + f^2} + B_3 \right),
\]
where \(B_1\), \(B_2\), and \(B_3\) are temperature- and pressure-dependent coefficients, and \(f_{rN}\), \(f_{rO}\) are the relaxation frequencies of nitrogen and oxygen. The expressions for these coefficients and relaxation frequencies are provided in the Supplementary Material. Constants such as the relative humidity, absolute temperature, and saturation vapor pressure are listed in Supplementary Material.

\subsection{Numerical Implementation}
The 3D Huygens' principle model and resulting acoustic streaming force were implemented in MATLAB. The forces (sliced in 2D plane) were exported as CSV files and imported into COMSOL Multiphysics 6.2 using interpolation to apply the forces as volume forces. A 2D mirror symmetry along the \(x = 0\) boundary and a laminar flow model was assumed for all models. The modeled domain had a width of \(200 \,~ \text{mm}\) and a height of \(400 \,~ \text{mm}\). Pressure outlet conditions were applied to the outer boundaries, while the bottom boundary (where the array was located) was modeled as a wall condition. Detailed descriptions of the numerical setup and mesh resolution are provided in the Supplementary Material. 

\subsection{Data Analysis}
The experimentally obtained flow velocity fields were further analyzed in MATLAB by applying a 4th-order Butterworth low-pass filter with a cutoff frequency of 10 Hz to remove high-frequency noise. To minimize edge effects during filtering, the time-series data for each spatial point in the velocity field was symmetrically padded (sample processing for randomly selected case study with original and filtered data is available in the Supplementary Material). Missing or non-finite values in the data were replaced using cubic spline interpolation to ensure continuity. After filtering, spline fitting was applied to smooth the data further. The maximum velocity was calculated at each frame across all spatial points, with the mean and standard deviation of these maxima computed over the recorded frames. A single sample was taken per condition, and the standard deviation reflects temporal variations in the acoustic streaming velocity, as discussed in the discussion section. This analysis was repeated for all experimental conditions, including variations in focal distances and cone angles. The results were subsequently compared to theoretical predictions from thermoviscous and atmospheric attenuation models to understand the most appropriate models for this type of scenario. 
\section{Results}
\label{sec:results}
\subsection{Qualitative Comparison between Experimental and Numerical Results}
Figure~\ref{fig:quiver_experiment_numerical} compares the streaming velocity fields obtained experimentally and numerically under different attenuation models. The experimental data in Figure ~\ref{fig:quiver_experiment_numerical}(a) were obtained using PIV and represent the time-averaged velocity field across all captured frames (focused beam at (0, 0, 80) mm, 5V). The results reveal a symmetric streaming flow with strong grating lobe jets angled outward from the transducer array, contributing significantly to the lateral flow structure. The central region shows moderate flow magnitudes, forming a focused jet along the axis.

Figure \ref{fig:quiver_experiment_numerical}(b) illustrates numerical results for a 2D mirror symmetry model with thermoviscous attenuation. This model significantly underestimates both the central flow intensity and the grating lobe contributions, resulting in a weaker overall flow. In Figure \ref{fig:quiver_experiment_numerical}(c), the 2D mirror symmetry model with atmospheric attenuation shows improvement, capturing the central flow distribution and magnitude more accurately. However, it still underestimates the grating lobe intensity, leading to a simplified lateral flow pattern.

\begin{figure*}[t]
    \centering
    \includegraphics[width=0.85\textwidth]{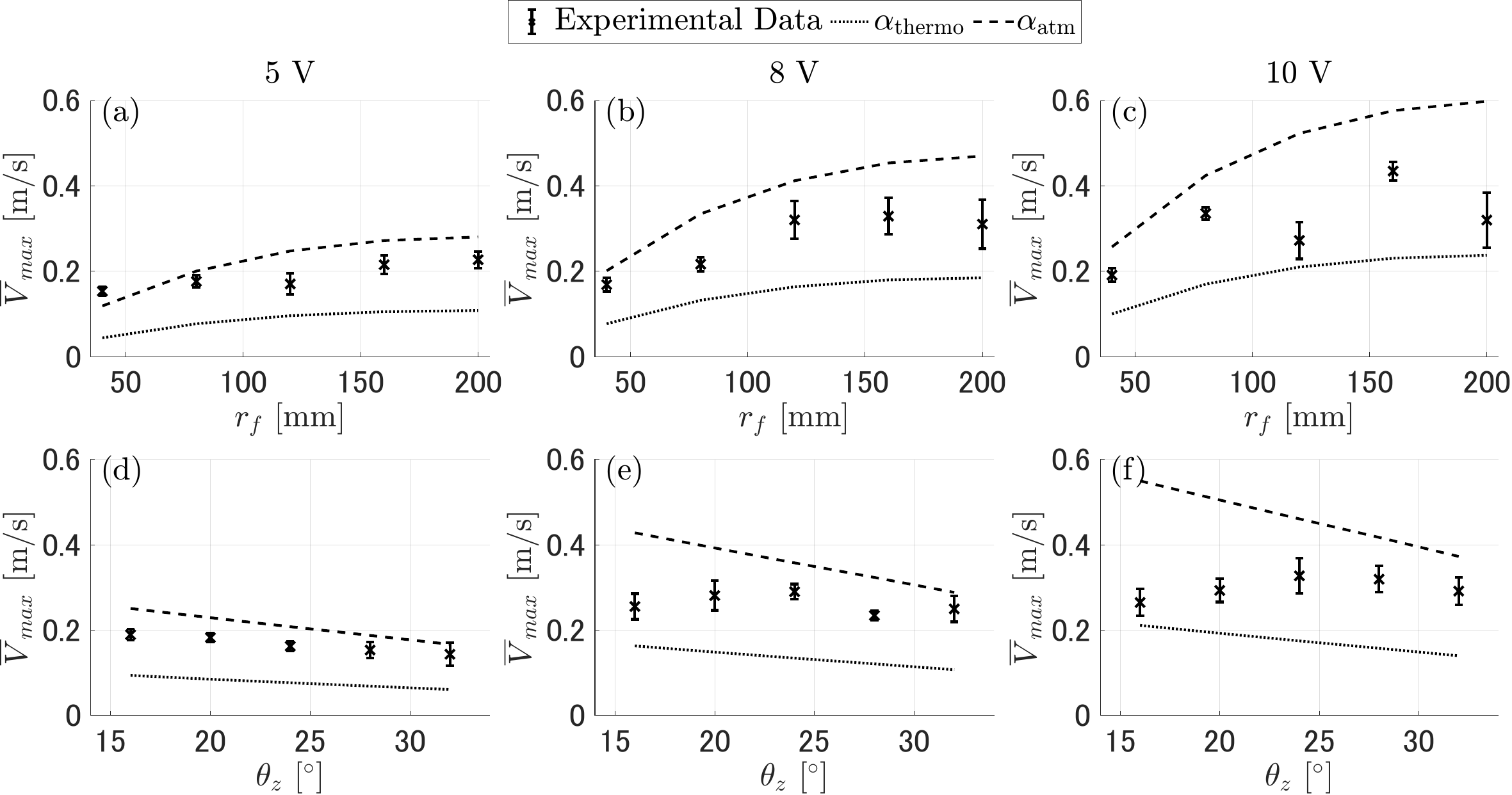}
    \caption{Comparison of experimental and theoretical maximum velocities for varying input voltages. (a-c) Mean maximum velocities ($\overline{V}{max}$) as a function of focal point ([0, 0, $r_f$] m) for input voltages of 5 V, 8 V, and 10 V, respectively. (d-f) Mean maximum velocities ($\overline{V}{max}$) as a function of cone angle ($\theta_z$) for input voltages of 5 V, 8 V, and 10 V, respectively. Experimental data are represented with error bars (standard deviation), while theoretical predictions are shown for thermoviscous (dotted lines) and atmospheric (dashed lines) attenuation models.}
    \label{fig:focal_bessel_comparison}
\end{figure*}

The comparison reveals that the 2D mirror symmetry model is capable of predicting the central streaming flow with reasonable accuracy, as evidenced by its ability to capture the strong central jet observed in the experimental data. However, the model cannot accurately replicate the lateral flow characteristics, particularly the prominent grating lobes. Despite this limitation, the model’s simplicity makes it a useful tool for modeling the central flow.
\subsection{Quantitative Comparison between Experimental and Numerical Results (2D Mirror Symmetry)}
Building on these insights, Figure~\ref{fig:focal_bessel_comparison} presents a quantitative comparison of the experimental and theoretical maximum velocities ($\overline{V}_{\mathrm{max}}$) under different attenuation models. In this analysis, we focus on the central flow velocity both in the experimental data and numerical simulations (2D mirror symmetry) to explore how the streaming velocity varies as a function of the focal distance ($r_f$) and cone angle ($\theta_z$) for single focal fields and Bessel beams, respectively.

Figure~\ref{fig:focal_bessel_comparison} presents a comparative analysis of the experimental and theoretical maximum velocities (\(\overline{V}_{\mathrm{max}}\)) for varying input voltages (5 V, 8 V, and 10 V) as functions of the focal distance (\(r_f\)) for single focal field and cone angle (\(\theta_z\)) for Bessel beams.

Figures~\ref{fig:focal_bessel_comparison}(a)–(c) illustrate the dependence of \(\overline{V}_{\mathrm{max}}\) on the focal distance for input voltages of 5 V, 8 V, and 10 V. The experimental data, represented with error bars indicating standard deviation, show a general increase in \(\overline{V}_{\mathrm{max}}\) with \(r_f\). This trend becomes more pronounced as the input voltage increases. At 5 V (Figure ~\ref{fig:focal_bessel_comparison}(a)), the experimental maximum velocity remains below 0.3~m/s across most focal distances, closely matching the atmospheric attenuation model (\(\alpha_{\mathrm{atm}}\)), except for that at \(r_f = 40\)~mm, where the experimental values slightly exceed the prediction of the thermoviscous attenuation model (\(\alpha_{\mathrm{thermo}}\)). At 8 V (Figure ~\ref{fig:focal_bessel_comparison}(b)), the experimental values exceed 0.3~m/s for larger focal distances. At 10 V (Figure ~\ref{fig:focal_bessel_comparison}(c)), the experimental \(\overline{V}_{\mathrm{max}}\) approaches 0.4~m/s, particularly at longer focal distances. The larger error bars in Figure~\ref{fig:focal_bessel_comparison}(c) reflect limitations of the measurement system, which operates at a frame rate of 240 FPS. Any particle displacements that exceed the system’s trackable range (as dictated by the per-frame displacement limit) were recorded as NaN and therefore excluded from the mean calculation, leading to an underestimated flow velocity. To capture these faster particle velocities accurately, an optical measurement system with a broader measurement range—such as a higher frame rate or alternative tracking method—would be needed.

Figures~\ref{fig:focal_bessel_comparison}(d)–(f) show the dependence of \(\overline{V}_{\mathrm{max}}\) on the cone angle for the Bessel beam. At 5 V (Figure ~\ref{fig:focal_bessel_comparison}(d)), the experimental maximum velocity remains below 0.2~m/s across all cone angles. At 8 V (Figure ~\ref{fig:focal_bessel_comparison}(e)), the experimental values approach 0.3~m/s. At 10 V (Figure ~\ref{fig:focal_bessel_comparison}(f)), the experimental data initially align closely with the thermoviscous attenuation model at smaller cone angles (\(\theta_z = 16^\circ\)), but gradually approach the predictions of the atmospheric attenuation model as the cone angle increases. 

Overall, the experimental results show good agreement with the numerical simulations and are bounded by the predictions of the thermoviscous and atmospheric attenuation models. Notable exceptions include \(r_f = 40\)~mm at 5 V in Figure ~\ref{fig:focal_bessel_comparison}(a), where the experimental data slightly exceed the thermoviscous attenuation model. Nevertheless, the overall consistency between experimental results and simulations at the central flow, reveals the predictive capability of the numerical model.

\section{Discussion}
\label{sec:discussion}
Building on these findings, we discuss the transient behavior of acoustic streaming, interaction between multi-focus fields and flow dynamics, and challenges associated with the experimental and numerical methodologies, aiming to provide a more comprehensive understanding of acoustic streaming generated by PATs.
\subsection{Temporal Response of Acoustic Streaming}
\begin{figure}
    \centering
    \includegraphics[width=0.4\textwidth]{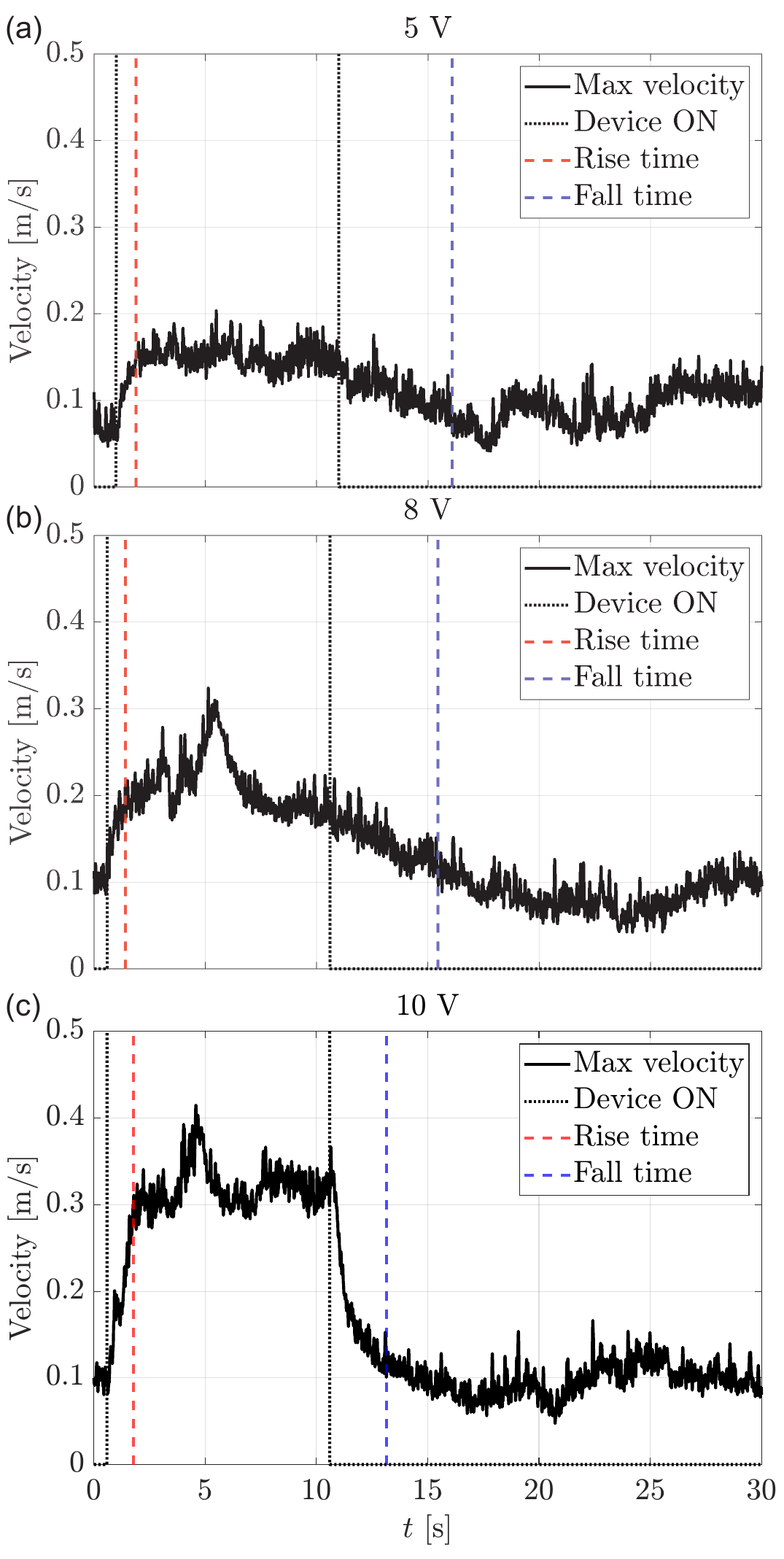}
    \caption{Time response of the streaming velocity for input voltages of 5 V (a), 8 V (b), and 10 V (c). The plots show the rise time, fall time, and steady-state velocity measured along the centerline at a focal distance of 80 mm.}
    \label{fig:time_response}
\end{figure}

In this section, we investigate the transient behavior of the flow field. Numerical simulations using a multi-physics finite element model (linear flow, stationary model) compute only the steady-state solution and cannot account for transient dynamics. However, since acoustic streaming is driven by the body force induced by the acoustic field, it does not develop immediately. The transient behavior of the acoustic streaming velocity was analyzed under consistent experimental conditions for three input voltages—5 V, 8 V, and 10 V—corresponding to Figs.~\ref{fig:time_response}(a), (b), and (c), respectively (see data statement for time lapse video of streaming field). In all cases, the focal distance was fixed at 80 mm with a singular focus configuration, and the velocity was measured along the centerline of the acoustic field. The plots display a 30 s time clip, during which the device was initially off, turned on for 10 s to record the streaming flow behavior, and subsequently turned off to observe the decay phase. Key metrics such as the mean velocity, maximum acceleration, maximum deceleration, rise time (time required to reach the mean velocity), and fall time (time required to return to the baseline after the device is turned off) were extracted to characterize the system's response.

At 5 V, the rise time is \(0.887\) s, with a mean velocity of \(0.149\) m/s and maximum acceleration of \(0.167\) m/s². The fall time is \(5.088\) s, with a maximum deceleration of \(-0.029\) m/s². At 8 V, the system responds faster, with a rise time of \(0.808\) s, mean velocity of \(0.209\) m/s, and maximum acceleration of \(0.259\) m/s². The fall time reduces to \(4.858\) s, and the maximum deceleration increases to \(-0.043\) m/s². At 10 V, the rise time increases slightly to \(1.192\) s, with a mean velocity of \(0.305\) m/s and maximum acceleration of \(0.256\) m/s². The fall time decreases significantly to \(2.550\) s, with a much higher maximum deceleration of \(-0.120\) m/s².

Across all cases, the rise time and acceleration during the device ON phase show a voltage-dependent trend. As the input voltage increases, the system achieves higher steady-state velocities. During the device OFF phase, the fall time and deceleration indicate faster dissipation of the streaming flow at higher voltages.

\subsection{Effect of Multi-Focus}
\begin{figure*}
    \centering
    \includegraphics[width=0.9\textwidth]{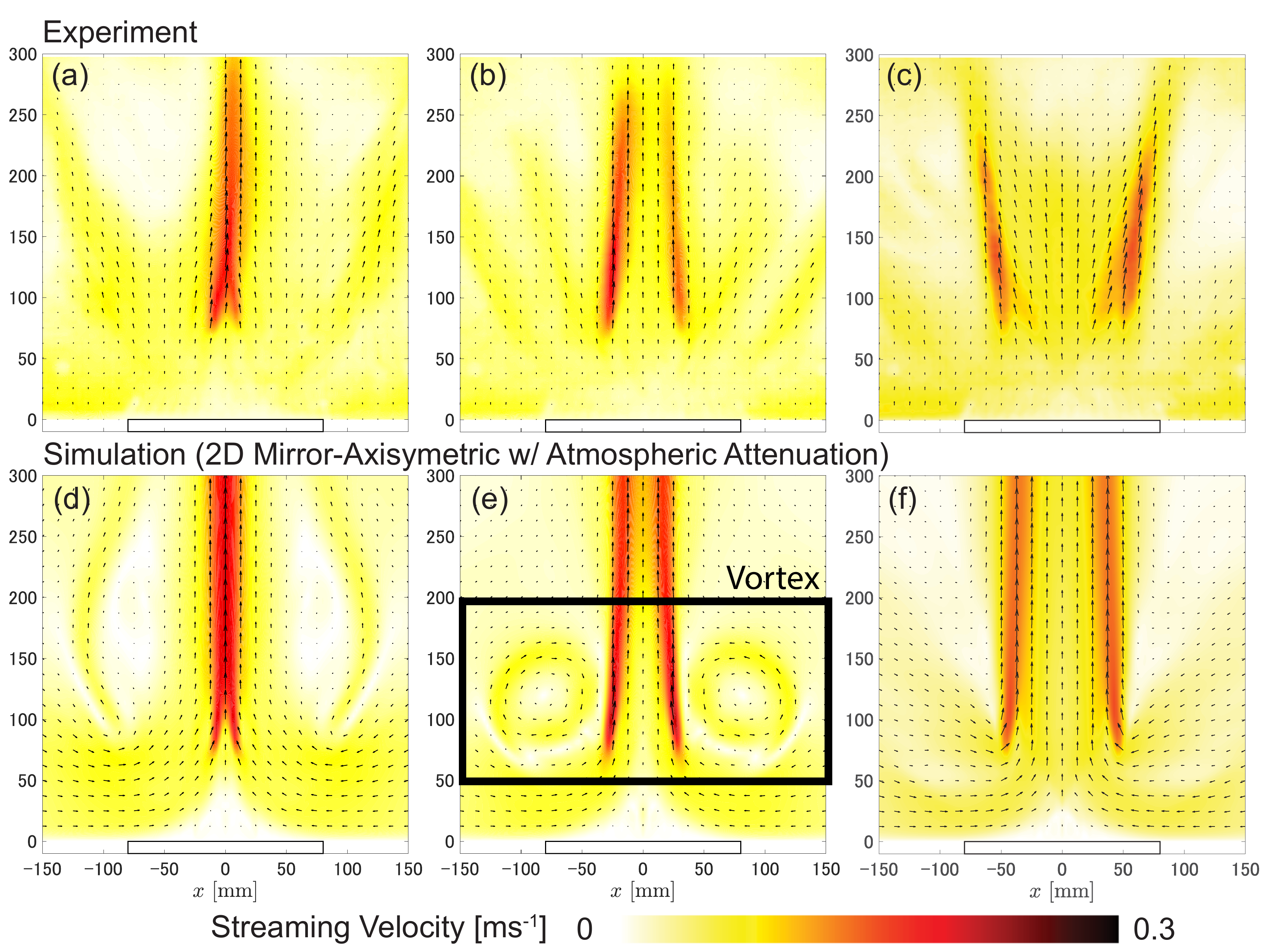}
    \caption{Comparison of experimental ((a)–(c)) and simulated ((d)–(f)) multi-focus streaming velocity fields at \(\pm 10\) mm (a, d), \(\pm 30\) mm (b, e), and \(\pm 50\) mm (c, f) along the \(z = 80\) mm plane. Simulations assume 2D line symmetry under atmospheric conditions.}
    \label{fig:multi_focus}
\end{figure*}

The capability to generate and control multiple focal points is essential in applications such as haptics and acoustic levitation, enabling simultaneous manipulation or interaction at distinct spatial locations. Understanding how multi-focus acoustic streaming fields differ from those generated by single, focused beams is critical for optimizing performance and ensuring accurate modeling in these advanced acoustic systems. The analysis of multi-focus acoustic streaming fields was conducted experimentally and numerically, with focal points placed at \((-10, 0, 80)\) mm and \((10, 0, 80)\) mm for Figure ~\ref{fig:multi_focus}(a) and (d), \((-30, 0, 80)\) mm and \((30, 0, 80)\) mm for Figure ~\ref{fig:multi_focus}(b) and (e), and \((-50, 0, 80)\) mm and \((50, 0, 80)\) mm for Figure ~\ref{fig:multi_focus}(c) and (f). The input voltage was fixed at 8 V for all cases. The numerical simulations were performed using a 2D mirror model at \(z = 0\), assuming atmospheric attenuation. The experimentally observed and simulated velocity fields were compared to evaluate the accuracy of the model and identify discrepancies.
For cases where the focal points are relatively close together (\(\pm 10\) mm in Figure ~\ref{fig:multi_focus}(a) and (d) and \(\pm 30\) mm in Figure ~\ref{fig:multi_focus}(b) and (e)), the experimental and numerical results show good agreement. Both the magnitude and spatial distribution of the streaming velocity fields match well, and the flow fields combine effectively into a single unified jet near the focal region. This agreement suggests that the symmetry and assumptions in the 2D model are well-suited for these configurations. However, as the focal points become farther apart (\(\pm 50\) mm in Figure ~\ref{fig:multi_focus}(c) and (f)), the simulation struggles to fully replicate the experimental results. In these cases, while the magnitude of the velocity field near the focal points is well-approximated, the direction of the flow shows notable discrepancies. For instance, in the experimental result Figure ~\ref{fig:multi_focus}(c), the flow is oriented outward from the focal points, whereas the simulation Figure ~\ref{fig:multi_focus}(f) predicts a straight upward flow. This mismatch is consistent with previous observations of directional errors in simulated side-lobe fields and highlights the need for improved attenuation models, better assumptions regarding weakly nonlinear acoustic systems, 3D models, and improved boundary conditions (e.g., accounting for imperfections in the wall conditions of the transducer array). Additionally, when the focal points are close together, the flow fields combine effectively, creating a unified structure. This behavior is evident in Figure ~\ref{fig:multi_focus}(a), (b), (d), and (e), where the flow is directed and cohesive. As the focal points become more separated, as in Figure ~\ref{fig:multi_focus}(c) and (f), the interaction between the flow fields diminishes, leading to distinct differences in the spatial orientation and coherence of the flow.

Another interesting observation is the formation of vortex fields in the simulation in Figure ~\ref{fig:multi_focus}(e) near the walls, likely caused by the interaction of the flow with the zero-velocity boundary condition. These vortex fields are not observed experimentally in Figure ~\ref{fig:multi_focus}(b), suggesting that the experimental system might not exhibit such strong interactions with the boundaries or that additional factors not captured in the simulation mitigate their formation.
\subsection{Limitations and Future Study}
The primary limitation of this study lies in the use of a two-dimensional model to simulate an inherently three-dimensional phenomenon. While the 2D mirror symmetry model captures the central flow reasonably well, extending the analysis to a full 3D simulation is expected to provide a more accurate representation of the observed behavior. The experimental results are well bounded between the thermoviscous and atmospheric conditions, offering a robust foundation for simulation. However, the assumed form of the streaming forces may warrant re-examination, as the current weakly nonlinear formulation does not perfectly fit with the experimental observations. 

Another limitation lies in the PIV experimental measurement setup. The current system reaches measurement limits, especially for faster flow fields. Upgrading optical equipment (i.e.~camera and laser equipment) will be essential for accurate measurement of higher velocities. Although the current laser sheet is sufficient to cover the transducer array, larger arrays pose challenges in ensuring adequate coverage and a wide range of view. 

These limitations underscore the importance of both improving the computational modeling approaches and advancing experimental measurement capabilities to better understand and characterize more complex acoustic streaming phenomena.
\section{Conclusions}
\label{sec:conclusions}
This study systematically investigated acoustic streaming fields generated by phased array transducers under various configurations, including single, Bessel and multi-focus setups, through a combination of experimental measurements and numerical simulations. The experimental results, validated against 2D mirror symmetry models, showed good agreement for cases with closely spaced multi focus, highlighting the efficacy of the current modeling approach. However, discrepancies in the flow direction and vortex formation were observed in cases with larger focal separations, emphasizing the need for refined attenuation models, nonlinear assumptions, 3D models, and boundary conditions. These results contribute to advancing the understanding and design of acoustic streaming for diverse applications, including haptics, cooling, and acoustic levitation.
\section*{Author Declaration}
The authors have no conflicts to disclose.
\section*{Author Contributions}
\textbf{Christopher Stone}: Conceptualization (equal); Data curation (lead); Formal analysis (equal); Methodology (equal); Software (equal); Visualization (equal); Writing – original draft (equal); Discussion (equal).  

\textbf{Yusuke Koroyasu}: Data curation (equal); Software (equal); Methodology (equal); Writing – review \& editing (equal); Discussion (equal).  

\textbf{Yoichi Ochiai}: Resources (equal); Methodology (equal); Writing – review \& editing (equal); Discussion (equal).  

\textbf{Akiko Kaneko}: Resources (equal); Methodology (equal); Writing – review \& editing (equal); Discussion (equal).  

\textbf{Bruce W. Drinkwater}: Conceptualization (equal); Methodology (equal); Funding acquisition (supporting); Project administration (equal); Supervision (equal); Writing – review \& editing (equal); Discussion (equal).  

\textbf{Tatsuki Fushimi}: Conceptualization (equal); Formal analysis (equal); Funding acquisition (lead); Methodology (equal); Project administration (equal); Resources (equal); Software (equal); Supervision (equal); Validation (lead); Visualization (equal); Writing – original draft (lead); Writing – review \& editing (equal); Discussion (equal). 

\section*{SUPPLEMENTARY MATERIAL}
See the supplementary material for overview of experimental setup, detailed implementation of the Huygens' principle model and phase calculation methods, attenuation coefficients, COMSOL multiphysics setup, and low pass filter

\section*{Data Statement}
The data that support the findings of this study are openly available in its supplementary material and figshare at
[]. (Currently private, and will be made public upon acceptance:  )

\section*{Acknowledement}
This work was supported by Japan Science and Technology Agency (JST) as part of Adopting Sustainable Partnerships for Innovative Research Ecosystem (ASPIRE), Grant Number JPMJAP2330. We would like to thank Murata Manufacturing Co., Ltd.~for providing transducers used in experiments.  
\nocite{*}
\bibliography{manuscript.bbl}

\end{document}